\documentclass[12pt]{iopart}

\usepackage{iopams}
\usepackage{soul}										
\setstcolor{red}	  
\usepackage{cite}

\usepackage{epsfig}
\usepackage{amsmath,amssymb}
\usepackage{bm}
\usepackage{graphicx}
\usepackage{cancel}
\usepackage{empheq}
\usepackage{fancyvrb}
\usepackage{ulem}
\usepackage{booktabs}
\usepackage{hyperref}
\usepackage{lineno}

\usepackage{color}
\usepackage{multirow}
\usepackage{makecell}
\usepackage{booktabs}
\usepackage{makecell}

\begin{document}


\title{The Non-thermal Energy Window for Laser-Driven Nuclear Reactions
}

\author{Eunseok Hwang}
\ead{hwangeunseok94@gmail.com}
\address{
    Department of Physics and OMEG Institute, Soongsil University, Seoul 06978, Republic of Korea
}

\author{Heamin Ko}
\ead{h.ko@gsi.de}
\address{
               Institut f\"ur Kernphysik, Technische Universit\"at Darmstadt, Schlossgartenstra{\ss}e 2, 64289 Darmstadt, Germany}

\author{Myung-Ki Cheoun}
\ead{cheoun@ssu.ac.kr}
\address{
    Department of Physics and OMEG Institute, Soongsil University, Seoul 06978, Republic of Korea
}

\author{Dukjae~Jang}
\ead{Corresponding Author: dukjaejang91@gmail.com}
\address{Department of Physics and OMEG Institute, Soongsil University, Seoul 06978, Republic of Korea}
\address{Department of Physics, Incheon National University, Incheon 22012, Republic of Korea}

\begin{abstract}
Laser-driven nuclear reactions proceed in non-equilibrium plasma conditions, producing ion energy distributions that are not Maxwellian. Nevertheless, fusion yields in such experiments are often interpreted using effective thermal descriptions based on the conventional Gamow window. In this work, we develop an analytical framework for evaluating nuclear reaction rates for non-thermal ions accelerated by the Target Normal Sheath Acceleration (TNSA) mechanism. Using a self-similar plasma expansion model, we derive a closed form expression for an effective reaction energy window and the corresponding fusion reactivity. The resulting effective energies differ systematically from those predicted by thermal models, indicating limitations of interpretations based on the conventional Gamow window in laser-driven environments. This framework provides a quantitative basis for analyzing fusion yields and for designing laser-driven nuclear experiments.
\end{abstract}

\date{\today}
\maketitle

\section{Introduction}
Laser-driven nuclear reactions provide access to fusion processes in plasma environments that are far from thermal equilibrium. In {\it pitcher-catcher} configurations employing ultra-intense, short-pulse lasers, ions are accelerated to multi-MeV energies with ultrashort durations and high instantaneous fluxes, enabling measurements of fusion yields under conditions distinct from those achieved with conventional accelerator-based approaches \cite{Wilks2001,Ditmire1999,Fuchs2006,Robson2007}. Such experiments involve non-thermal ion populations, dense plasmas, and rapidly evolving interaction geometries, and have been applied to studies of nuclear reactions in plasma environments \cite{Barbui2013,Storm2013,Willingale2011,Negoita2016,Daido2012,Macchi2013,Passoni2010,TerAvetisyan2005}.

In laser-driven nuclear experiments, fusion yields are interpreted through nuclear reaction rates. Conventionally, fusion yields and effective reaction energies are inferred using thermal formalisms based on Maxwell–Boltzmann ion distributions and the conventional Gamow window \cite{Angulo1999,Arnould2003,Iliadis2015,Rauscher2010}. While this approach is applicable to thermally equilibrated plasmas, ions accelerated by Target Normal Sheath Acceleration (TNSA) exhibit non-Maxwellian energy spectra with extended high-energy components \cite{Mora2003,Robson2007,Passoni2010,Macchi2013}. In addition, the short interaction time in laser-driven experiments limits the establishment of thermal equilibrium. These features indicate that reaction energies inferred using thermal assumptions may not accurately represent the dominant energy range in laser-driven environments.

In this Letter, we present for the first time an analytical derivation of an effective reaction-energy window for a non-Maxwellian ion distribution generated by the TNSA mechanism. We adopt the self-similar solution derived from the plasma expansion model, which provides a physically motivated description of ion energy distributions in laser-driven experiments \cite{Mora2003,Fuchs2006}. This distribution is combined with the nuclear reaction cross section, expressed in terms of the astrophysical $S$-factor and the Coulomb penetration factor, allowing the reaction-rate integral to be formulated explicitly. Within this formulation, we derive closed-form expressions for the effective reaction energy and the corresponding fusion reactivity for both non-resonant and resonant processes. The resulting framework provides a quantitative basis for identifying the dominant reaction-energy window in laser-driven nuclear reactions. Furthermore, we discuss how to extract low-energy nuclear cross sections and astrophysical $S$-factors from laser-driven fusion yields, with relevance to both nuclear fusion studies and laboratory nuclear astrophysics.

\section{Formalism}
The averaged fusion reactivity for beam ($b$) and target ($t$) ions is given by
\begin{eqnarray}
\langle \sigma v \rangle &=& \int\!\int f_b(\mathbf{v}_b) f_t(\mathbf{v}_t) \sigma(E) v\,d\mathbf{v}_b\,d\mathbf{v}_t, 
\label{eq:react_rate}
\end{eqnarray}
where $\sigma(E)$ is the reaction cross section as a function of the center-of-mass (CM) energy $E$, $v = |\mathbf{v}_b - \mathbf{v}_t|$ is the relative velocity, and $f_b(\mathbf{v}_b)$ and $f_t(\mathbf{v}_t)$ are the normalized velocity distribution functions of beam and target ions, respectively.

\begin{figure}[t]
\centering
\includegraphics[width=1\linewidth]{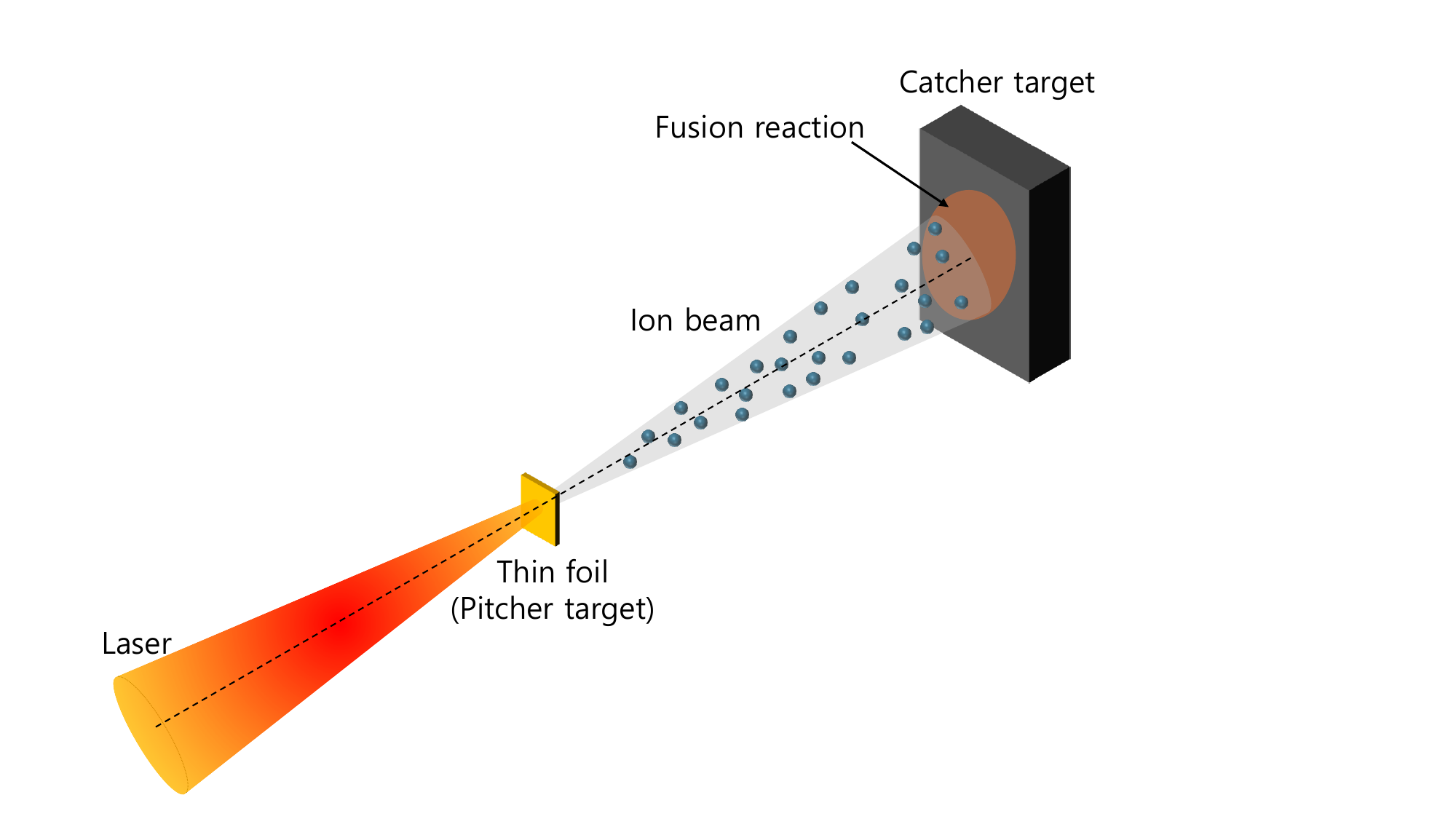}
\caption{Schematic of the pitcher-catcher experimental setup. A high-intensity laser pulse irradiates a thin foil target (pitcher target), accelerating a beam of ions, which is then incident on a secondary target (catcher target). Nuclear reactions occur as the laser-accelerated ions interact with the secondary target.}
\label{fig:exp_setup}
\end{figure}

Our focus is to derive an analytical form of Eq.\,(\ref{eq:react_rate}) on the {\it pitcher-catcher} configuration illustrated schematically in Fig.~\ref{fig:exp_setup}. In this setup, a short-pulse, high-intensity laser irradiates a thin foil, expelling relativistic electrons and creating a strong electrostatic sheath field on its rear surface. This sheath field accelerates ions normal to the target surface via TNSA. The effective sheath area is given as:
\begin{eqnarray}
S_{\rm sheath} = \pi \left(r_0 + d_t \tan\theta\right)^2,
\label{sheath}
\end{eqnarray}
where $r_0$ is the laser focal spot radius, $d_t$ is the foil thickness, and $\theta$ is the divergence half-angle of the hot electron cloud \cite{Fuchs2006}. The accelerated ion beam is subsequently incident on a secondary catcher target positioned on the beam axis.

In the sheath, the hot-electron distribution is assumed to be a Maxwell–Boltzmann distribution, so the electron density reads:
\begin{eqnarray}
n_e(x,t) = n_{e0}\,\exp\!\Big[\frac{e\Phi(x,t)}{k_B T_e}\Big]~,
\label{ne}
\end{eqnarray}
where $n_{e0}$ is the initial electron density at the target rear, $\Phi(x,t)$ is the electrostatic potential, $k_B$ is the Boltzmann constant, and $T_e$ is the effective electron temperature. We adopt $T_e$ as the ponderomotive scaling for relativistic laser plasma interactions \cite{PhysRevLett.69.1383}:
\begin{eqnarray}
k_BT_e \;=\; m_e c^2\! \left[ \sqrt{1 + \frac{I\,\lambda_\mu^2}{1.37 \times 10^{18}}}\,-\,1 \right]~,
\label{Te}
\end{eqnarray}
where $m_e$ is the electron mass, $c$ is the speed of light, $I$ is the peak laser intensity in ${\rm W\,cm}^{-2}$, and $\lambda_\mu$ is the laser wavelength in $\mu$m. The initial density of the number of hot electrons can be written as $n_{e0} = N_e/(c\,\tau_{\rm laser}\,S_{\rm sheath})$, where $\tau_{\rm laser}$ is the duration of the laser pulse and $N_e$ is the total number of hot electrons. 
$N_e$ is given by $N_e = f\,E_{\rm laser}/(k_B T_e)$, with the laser pulse energy $E_{\rm laser}$ and the laser-to-hot-electron energy conversion efficiency $f$. We adopt the typical value $f \approx 1.2 \times 10^{-15} I^{0.74}$ based on empirical fits \cite{10.1063/1.872867, PhysRevE.56.4608}.

The plasma expansion is governed by the electric potential generated by the electron sheath. In the one-dimensional model adopted in this Letter, the plasma expansion is described by Poisson's equation, coupled with the ion continuity and momentum equations:
\begin{eqnarray}
&& \epsilon_0\,\frac{\partial^2 \Phi}{\partial x^2} =  e(n_e - Z_i n_i)\,,
\label{Poisson}\\[6pt]
&&\frac{\partial n_i}{\partial t} + \frac{\partial}{\partial x}(n_i v_i) = 0\,,
\label{eq:con}\\[6pt]
&&\frac{\partial v_i}{\partial t} + v_i\frac{\partial v_i}{\partial x} = -\,\frac{Z_i e}{m_i}\frac{\partial \Phi}{\partial x}\,,
\label{eq:eom}
\end{eqnarray}
where $n_i(x,t)$ and $v_i(x,t)$ denote the ion number density and velocity, respectively, $Z_i$ is the charge number of ion, $m_i$ is the ion mass, and $\epsilon_0$ is the vacuum permittivity. The initial conditions assume that at $t=0$, ions are initially at rest and confined to the region $x \le 0$, while electrons immediately form a Boltzmann sheath extending into the region $x > 0$, following Eq.\,(\ref{ne}). Solving the coupled system of equations (\ref{Poisson})–(\ref{eq:eom}) with these initial conditions yields the energy spectrum of ions accelerated from the rear surface of the target.

Assuming the plasma expansion remains quasi-neutral ($n_e \approx Z_i n_i$) over the acceleration time $t_{\rm acc}$, an analytic self-similar solution for the ion spectrum can be derived \cite{Mora2003}. In the TNSA regime, this solution predicts the number of ions per unit energy as \cite{Mora2003, Fuchs2006} 
\begin{eqnarray}
\frac{dN_i}{dE_i} \;=\; \frac{n_{e0}\, c_{si} \,t_{\rm acc}\,S_{\rm sheath}}{\sqrt{2 Z_i E_i\,k_B T_e}}\;\exp\!\left[-\,\sqrt{\frac{2 E_i}{Z_i k_B T_e}}\right]~,
\label{N_ss}
\end{eqnarray}
where $ c_{si} = \sqrt{Z_i k_B T_e/m_i}$ is the ion sound speed and $Z_i$ is the charge number of ion beam. The $t_{\rm acc}$ is typically related to the laser pulse duration $\tau_{\rm laser}$ and can be adopted as $t_{\rm acc} \approx 1.3\,\tau_{\rm laser}$. From the self-similar solution in Eq.\,(\ref{N_ss}), the corresponding normalized energy distribution for the ion beam is defined as
\begin{eqnarray}
f_{i,ss}(E_i)  \equiv \frac{1}{N_{0,i}} \frac{dN_i}{dE_i} = \frac{1}{\sqrt{2 Z_i E_i\,k_B T_e}}\;\exp\!\left[-\,\sqrt{\frac{2 E_i}{Z_i k_B T_e}}\right],
\label{f_ss}
\end{eqnarray}
where $N_{0,i} = n_{e0}\,  c_{si} \,t_{\rm acc}\,S_{\rm sheath}$ is the total number of accelerated ions of species $i$. We note that the non-thermal distribution $f_{i,ss}(E_i)$ depends on the electron temperature $T_e$, not on the ion temperature.

\begin{figure}[t]
\centering
\includegraphics[width=0.75\linewidth]{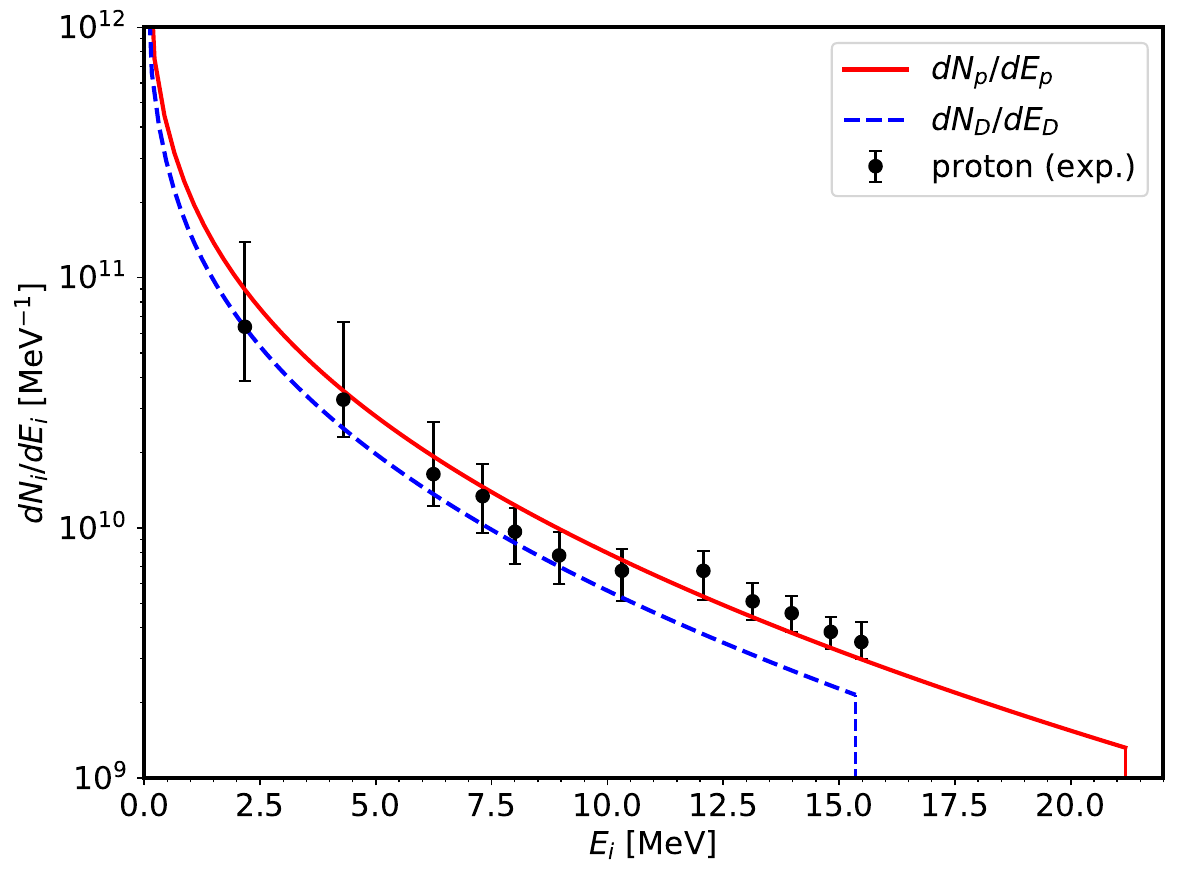}
\caption{Energy spectra of protons (red solid line) and deuterons (blue dashed line) calculated from a self-similar solution, compared with the experimental proton-beam data (black dots) \cite{Fuchs_PRL}. We adopt a laser intensity of $I = 3 \times 10^{19}\,\mathrm{W/cm^2}$ and a wavelength of $\lambda = 1.057\,\mu\mathrm{m}$, consistent with the 100-TW laser experiment at LULI \cite{Fuchs_PRL}. Under these parameters, corresponding to $k_B T_e = 2.068\,\mathrm{MeV}$, the acceleration time satisfies $\omega_{pi}t_{acc} = 11.08$ for protons and $\omega_{pi}t_{acc} = 7.839$ for deuterons. The target thickness is taken to be $20\,{\rm \mu m}$ for an aluminum target, following Ref.\,\cite{Fuchs_PRL}, and is used to determine the maximum energy of the accelerated ions.} 
\label{fig:compare_exp}
\end{figure}

Figure \ref{fig:compare_exp} shows the differential ion spectra $dN_i/dE_i$ for proton and deuteron beams calculated from Eq.(\ref{N_ss}). This non-thermal distribution deviates from the Maxwell–Boltzmann form. As shown in comparison with the proton-beam data from Ref.\,\cite{Fuchs_PRL}, the self-similar solution exhibits good agreement with the experiment. The analytic solution, however, is valid only when the initial Debye length $\lambda_{D0}$ is much smaller than the plasma expansion scale, i.e., when $\omega_{pi} t_{\rm acc} \gg 1$, where $\omega_{pi} = \sqrt{n_{e0}  Z_i e^2/(m_i \epsilon_0)}$ is the ion plasma frequency. At extremely high intensities and ultrashort pulse durations that violate this condition, the quasi-neutral approximation breaks down, and the ion spectrum can be computed by numerically solving Eqs.\,(\ref{Poisson})–(\ref{eq:eom}). Alternatively, more rigorous theoretical frameworks, such as quasi-static models based on a self-consistent solution of the Poisson equation, can be employed \cite{PEREGO201189, perego_target_2012, Passoni_2013}. In this work, we restrict our analysis to the regime in which Eq.(\ref{f_ss}) remains valid.

We also note that the pitcher-target thickness enters the present calculation only through the maximum ion energy in the present work. In the plasma-expansion model, increasing the target thickness reduces the maximum ion energy \cite{Mora2003}, and this trend is broadly consistent with the experimental data \cite{Fuchs2006}. In realistic laser–target interactions, however, the ion-energy distribution also depends on the target composition, the laser-generated preplasma, and the impurity profile at the rear surface. A quantitative treatment of these effects requires more detailed particle-in-cell (PIC) simulations.

Identifying the ion distribution in Eq.\,(\ref{f_ss}) with the beam distribution in Eq.\,(\ref{eq:react_rate}), i.e., $f_{b}(E_b) = f_{i,ss}(E_i)$, and noting that the ions in the secondary (catcher) target are much slower velocities than the ion beam, i.e., $v \simeq |\mathbf v_i|$, the normalization $\int f_t(\mathbf v_t)\,\mathrm d\mathbf v_t = 1$ allows Eq.\,(\ref{eq:react_rate}) to be rewritten as
\begin{eqnarray}\label{eq:rate_beam}
\langle \sigma v \rangle
&\simeq& \int f_{i,\mathrm{ss}}(\mathbf v_i)\,\sigma(E)\,|\mathbf v_i|\,\mathrm d\mathbf v_i \\[6pt] \nonumber
&=& \sqrt{\frac{2}{m_i}} \int_0^\infty\sqrt{E_i}\, f_{i,ss}(E_i)\,\sigma(E)\,dE_i~,
\end{eqnarray}
where $E=\mu v^2/2 \simeq \mu v_i^2/2 = \mu/m_i E_i$ with $E_i= m_i v_i^2/2$, and the reduced mass is $\mu= m_i m_t/(m_i+m_t)$.

For non-resonant reactions, the cross section can be written in terms of the astrophysical $S$-factor $S(E)$ and the penetration factor for the Coulomb barrier $P(E)$:
\begin{equation}
\sigma(E)=\frac{S(E)}{E}\,P(E),
\label{sigma}
\end{equation}
where $P(E)=\exp[-2\pi\eta(E)]$ with the Sommerfeld parameter
$\eta(E)=Z_i Z_t e^2/(\hbar v)$. Equivalently, defining the Gamow energy $E_G = 2\mu c^2\left(\pi \alpha Z_i Z_t \right)^2$, one has $2\pi\eta(E)=\sqrt{E_G/E}$. Substituting Eq.\,(\ref{sigma}) into Eq.\,(\ref{eq:rate_beam}), we obtain an explicit form of the integrand in Eq.\,(\ref{eq:rate_beam}). Defining this integrand as $I(E_i)$, one finds that $I(E_i)$ is proportional to
\begin{equation}
I(E_i)\;\propto\;\frac{\sqrt{E_i}}{E}\,f_{i,\mathrm{ss}}(E_i)\,
\exp\!\left[-\sqrt{\frac{E_G}{E}}\right],
\label{integrand}
\end{equation}
where $S(E)$ is taken outside the integral and evaluated at the effective energy $E_0$, assuming it varies slowly with energy. The $E_0$, analagous to the procedure used to obtain the Gamow peak energy in thermal plasma, is determined from $\left.\mathrm d I(E_i)/\mathrm d E_i\right|_{E_i=E_0}=0$, from which one obtains
\begin{equation}
E_{0} \;=\; \frac{Z_i k_B T_e}{8}\left[\sqrt{\,4 + \sqrt{\frac{32\,m_i\,E_G}{\mu\,Z_i k_B T_e}}}\;-\;2\right]^{\!2}.
\label{E0}
\end{equation}
This effective energy $E_0$ serves as the non-thermal analogue of the Gamow peak, corresponding to the energy at which the fusion reactivity is maximized for a self-similar ion distribution produced by high-intensity laser acceleration.

\section{Results}
\begin{figure}[t]
\centering
\includegraphics[width=0.75\linewidth]{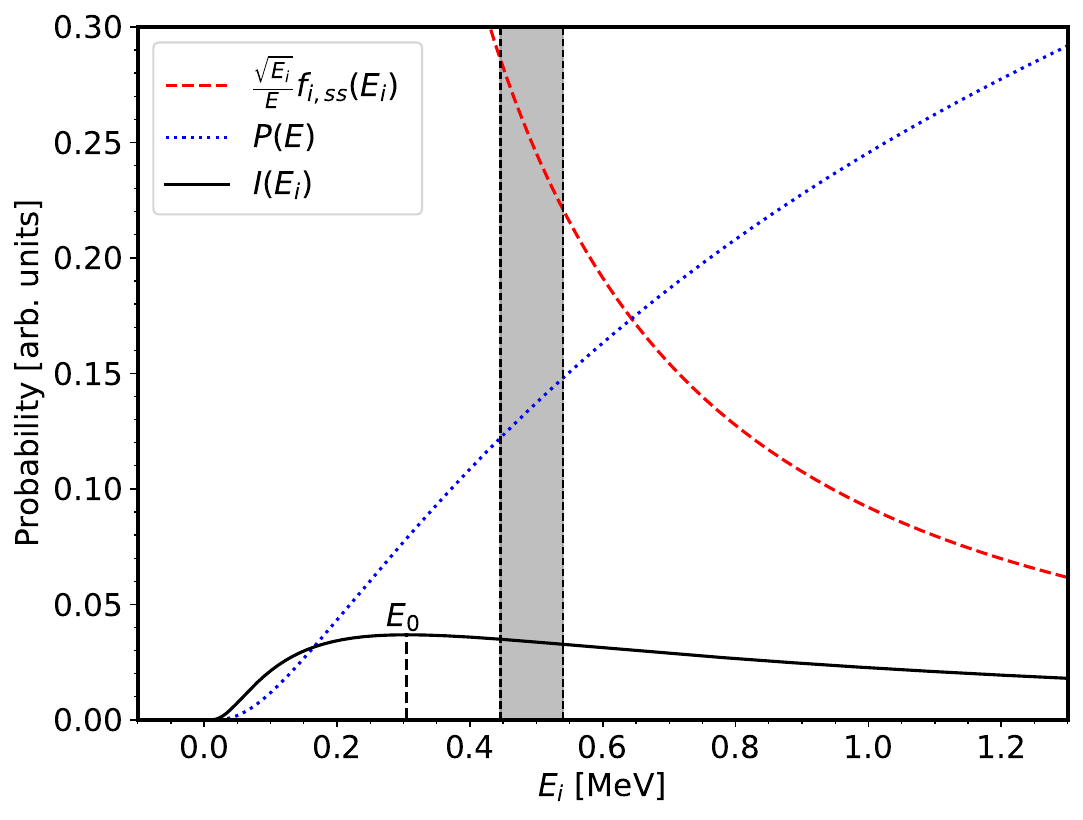}
\caption{
$I(E_i)$ (black solid line), $(\sqrt{E_i}/E)f_{i, ss}(E_i)$ (red dashed line), and $P(E)$ (blue dotted line) as a function of $E_i$ for the D+D reaction. 
For this calculation, the electron temperature is fixed as $k_B T_e = 2.068\,\mathrm{MeV}$, consistent with Fig.\,\ref{fig:compare_exp}. The peak energy is found to be $E_i  = 0.305\,\mathrm{MeV} \,(\equiv E_0)$. The shaded region at $E_i =0.494^{+0.046}_{-0.048}\,{\rm MeV} \,(\equiv E_{0,G})$ denotes the conventional Gamow peak energy in a thermal plasma, calculated using $k_BT_{\rm eff,D} = 0.7 \pm 0.1 \text{ MeV}$, where $k_BT_{\rm eff,D} =0.7\,{\rm MeV}$ is the best-fit value reported in Ref.\,\cite{Fuchs_PRL}.
}
\label{fig:distribution}
\end{figure}

For the D+D reaction, Fig.~\ref{fig:distribution} shows the $I(E_i)$, $P(E)$, and $(\sqrt{E_i}/E)f_{i, ss}(E_i)$ as a function of the $E_i$ for the same condition in Fig.\,\ref{fig:compare_exp}. The integrand exhibits a maximum at $E_i=E_0$, determined by Eq.\,(\ref{E0}), which defines the effective reaction energy, that is, the energy range dominating the contribution to $\langle\sigma v\rangle$. 
On the other hand, the conventional Gamow peak energy can be estimated using an effective ion temperature obtained by fitting the ion energy spectrum. Under the LULI conditions adopted in our calculation, the deuteron effective temperature is reported to be $k_B T_{\rm eff,D} = 0.7\,\mathrm{MeV}$ \cite{Fuchs_PRL}, which corresponds to a conventional Gamow peak energy of $E_{0, G} = 0.494\,\mathrm{MeV}$ for the D + D reaction. This value is approximately 1.6 times higher than our predicted peak energy, $E_0 = 0.305\,\mathrm{MeV}$, indicating that the effective reaction energy in our non-thermal model lies below the energy range predicted by the conventional Gamow window.

Using the self–similar ion beam distribution, the reactivity $\langle \sigma v \rangle$ in Eq.\,(\ref{eq:rate_beam}) can be evaluated in a closed form. Under the assumption that the astrophysical $S$-factor varies slowly with energy, we approximate $S(E)\simeq S(E_0)$ at the effective energy $E_0$ obtained above. Taking $S(E_0)$ outside the integral and inserting $P(E)=\exp[-\sqrt{E_G/E}]$ with the $f_{i,\mathrm{ss}}(E_i)$ into Eq.\,(\ref{eq:rate_beam}), we obtain
\begin{eqnarray} \label{eq_an_reac} 
\left\langle \sigma v \right\rangle &\simeq& \frac{S(E_0)}{\mu } \sqrt{\frac{m_i}{Z_i k_B T_e}} \\ \nonumber
&\times& \int_0^\infty \frac{1}{E_i} \exp \left[-\sqrt{\frac{2 E_i}{Z_i k_B T_e}} - \sqrt{\frac{m_i}{\mu} \frac{E_G}{E_i}}\right] dE_i~ \\[12pt] \nonumber
&=& \frac{4 S(E_0)}{\mu}\sqrt{\frac{m_i}{Z_i k_B T_e}} K_0 \left(\left[\frac{32 m_i E_G}{\mu Z_i k_B T_e}\right]^{\frac{1}{4}}\right),
\end{eqnarray}
where $K_{0}$ is the modified Bessel function of the second kind of order zero. In realistic conditions, the ion distribution exhibits a finite cutoff energy given by $ E_{\mathrm{max},i} = 2 Z_i k_B T_e \left[\ln\!\left( \tau_{pi} + \sqrt{ \tau_{pi}}^2+1\right)\right]^2$,
where $ \tau_{pi} = \omega_{pi} t_{\rm acc}/\sqrt{2e}$ \cite{Mora2003, Fuchs2006}. Hence, the upper limit of the integral in Eq.\,(\ref{eq_an_reac}) should, in principle, be taken to be $ E_{\mathrm{max},i}$ rather than infinity. However, when $\omega_{pi} t_{\rm acc} \gg 1$ or $I\lambda^{2}$ is sufficiently large, the cutoff energy becomes high enough that it can be treated as effectively infinite, in which case Eq.\,(\ref{eq_an_reac}) remains applicable.

The analytic expression in Eq.\,(\ref{eq_an_reac}) provides an approximate form of the reactivity, 
highlighting its dependence on the reduced mass $\mu$, the electron temperature $T_e$, 
the Gamow energy $E_G$, and the astrophysical $S$-factor $S(E_0)$. Consequently, once the target composition, beam properties, laser intensity, and wavelength are specified, the dominant energy region for the reaction is determined by Eq.\,(\ref{E0}), 
and the fusion reactivity can then be evaluated using Eq.\,(\ref{eq_an_reac}).

In Eq.\,(\ref{eq_an_reac}), there exists an optimal $T_e$ that maximizes the reactivity $\langle \sigma v \rangle$, denoted by $T_{e,\mathrm{max}}$. 
This value is obtained from the condition $d \langle \sigma v \rangle/d (k_B T_e)|_{k_B T_e = k_B T_{e,\mathrm{max}}}=0$,
which yields
\begin{equation}
k_B T_{e,\mathrm{max}} \simeq 0.1720\,B,
\label{Tmax}
\end{equation}
where $B = 32\,m_i E_G / (\mu Z_i)$. 
At this temperature, the maximum reactivity is given by
\begin{equation}
\langle \sigma v \rangle_{\mathrm{max}} 
\simeq S(E_0)\left[0.4815\,\sqrt{\frac{1}{2\mu E_G}}\right].
\label{sigmav_max}
\end{equation}
This result indicates that $\langle \sigma v \rangle$ attains a finite maximum in laser-driven environments. Physically, this maximum results from the competition between the Coulomb-penetration factor and the ion-energy distribution. As $I\lambda^2$ increases, the extended high-energy ion tail initially enhances the reactivity. At sufficiently high intensity, however, the normalization and broadening of the ion distribution reduce its magnitude in the effective reaction-energy region, leading to the finite maximum in Eq.~(\ref{sigmav_max}).

\begin{figure}[t]
\centering
\includegraphics[width=0.75\linewidth]{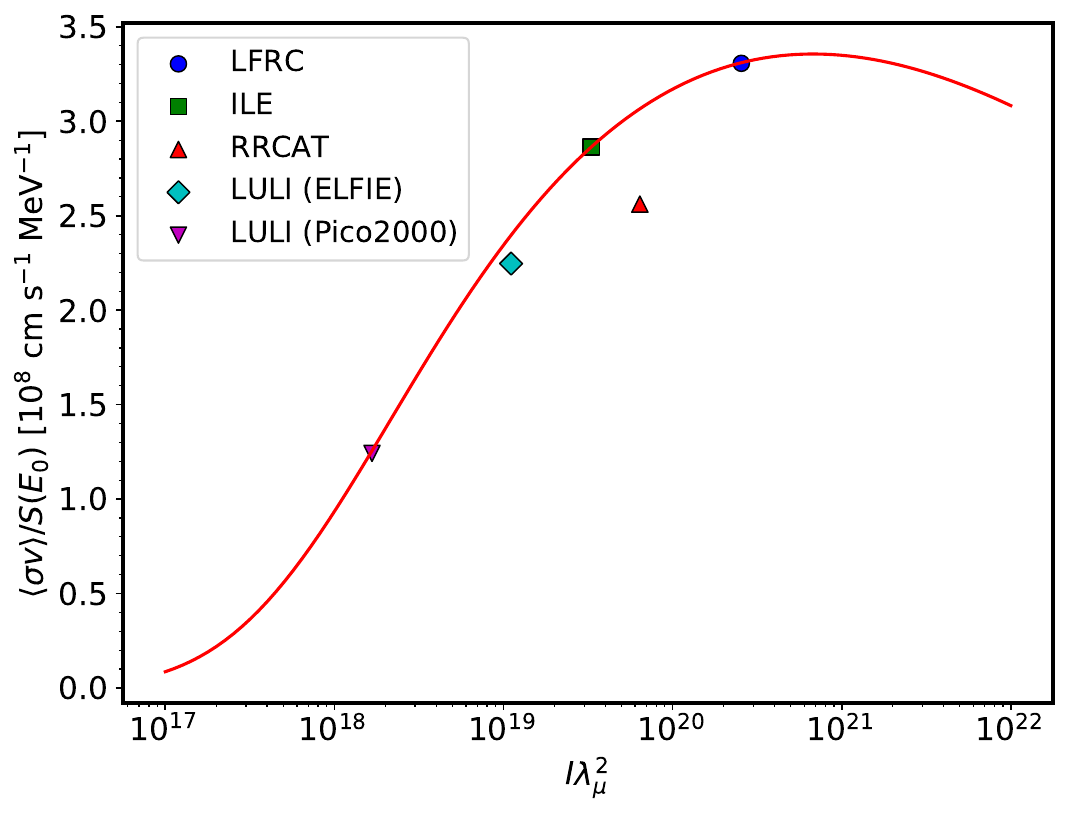}
\caption{
Ratio of the reactivity to the astrophysical $S$-factor at the peak energy, $\langle \sigma v \rangle / S(E_0)$, for the D+D reaction. The red solid curve is obtained from the analytical expression in Eq.~\eqref{eq_an_reac}. The maximum occurs at $k_B T_e = 10.85\,\mathrm{MeV}$, corresponding to $I \lambda_\mu^{2} = 6.752 \times 10^{20}$, where $\langle \sigma v \rangle / S(E_0) = 3.356 \times 10^{8}\,\mathrm{cm\,s^{-1}\,MeV^{-1}}$. Each marker denotes the reactivity obtained from numerical integration using the ion distribution function numerically calculated from the coupled Eqs.\,(\ref{Poisson})--(\ref{eq:eom}) with the corresponding $\omega_{pi} t_{\rm acc}$ values for different experimental conditions: XG-III at LFRC (blue circles), LFEX at ILE (green squares), 150-TW Ti:Sa laser at RRCAT (red triangles), ELFIE laser at LULI (cyan diamonds), and Pico2000 at LULI (magenta inverted triangles). The values corresponding to the markers are obtained using only the laser parameters reported for each experimental facility, which are listed in Table~\ref{tab:dd_laser_params}.
}
\label{fig:sigmav}
\end{figure}

For the D+D reaction, Fig.\,\ref{fig:sigmav} shows the ratio of the reactivity to $S(E_0)$, 
$\langle \sigma v \rangle / S(E_0)$, as a function of $I\lambda_\mu^{2}$. The maximum reactivity of this reaction is numerically predicted at $k_B T_e = 10.85\,{\rm MeV}$, 
corresponding to $I\lambda_\mu^{2} = 6.752\times10^{20}$, in agreement with Eq.\,(\ref{Tmax}). In Fig.\,\ref{fig:sigmav}, we compare the analytical expression with numerical evaluations of the D+D reactivity for laser parameters reported in several previous high-power-laser experiments\footnote{The original purpose of these experiments was to study the $p + {}^{11}\mathrm{B}$ reaction. We use only their reported laser parameters to evaluate the D + D reactivity for the corresponding laser configurations.}: the XG-III laser at the Laser Fusion Research Center (LFRC) \cite{wei2023(XG-III)}, the LFEX laser at the Institute of Laser Engineering (ILE) \cite{10.3389/fphy.2020.00343}, the 150-TW Ti:Sa laser at the Raja Ramanna Centre for Advanced Technology (RRCAT) \cite{Tayyab_2019}, the ELFIE laser at LULI \cite{Baccou_Depierreux_2015}, and the Pico2000 laser at LULI \cite{labaune_fusion_2013}. The corresponding laser parameters and evaluated quantities used in this comparison are summarized in Table~\ref{tab:dd_laser_params}.

Except for the 50-TW Ti:Sa laser condition at RRCAT, the analytical expression shows good agreement with the numerical results. In the RRCAT case, the condition $\omega_{pi} t_{\rm acc} = 2.74$ and $I \lambda_\mu^2 = 6.4 \times 10^{19}$ implies that the cutoff energy is not sufficiently large for the infinite-energy approximation to hold. Consequently, the expression in Eq.\,(\ref{eq_an_reac}) becomes inadequate and requires numerical evaluation over the relevant energy range. More generally, when $\omega_{pi}t_{\rm acc}$ is not sufficiently larger than unity, the self-similar approximation itself would lose quantitative validity. This limitation can become important for high-intensity configurations involving ultrashort pulses, for which the acceleration time could be too short to satisfy $\omega_{pi}t_{\rm acc}\gg1$.

\begin{table}[t]
\centering
\caption{Laser parameters reported in previous laser-driven $p+{}^{11}{\rm B}$ fusion experiments. Using only the reported laser parameters ($E_{\rm laser}$, $I$, $\tau_{\rm laser}$, $\lambda_\mu$, and $d_t$), rather than the measured fusion yields, we evaluated $k_B T_e$, $\omega_{pi}t_{\rm acc}$, and $\langle\sigma v\rangle/S(E_0)$ for the corresponding D+D configurations.}
\label{tab:dd_laser_params}
\footnotesize
\setlength{\tabcolsep}{3pt}
\renewcommand{\arraystretch}{0.95}
\begin{tabular}{@{}llcccccccc@{}}
\toprule
 &  & \multicolumn{5}{c}{Experimental Parameters}
 & \multicolumn{3}{c}{Evaluated Quantities} \\
\cmidrule(lr){3-7} \cmidrule(l){8-10}
Where 
& Laser 
& \makecell{$E_{\rm laser}$\\(J)}
& \makecell{$I$\\(W cm$^{-2}$)}
& \makecell{$\tau_{\rm laser}$\\(fs)}
& \makecell{$\lambda_L$\\($\mu$m)}
& \makecell{$d_t$\\($\mu$m)}
& \makecell{$k_B T_e$\\(MeV)}
& $\omega_{pi}t_{\rm acc}$
& \makecell{$\langle\sigma v\rangle/S(E_0)$\\($10^8$ cm s$^{-1}$ MeV$^{-1}$)} \\
\midrule
LFRC \cite{wei2023(XG-III)}
& XG-III
& 120
& $2.3\times10^{20}$
& 800
& 1.053
& 10
& 6.48
& 23.5
& 3.30 \\
ILE \cite{10.3389/fphy.2020.00343}
& LFEX
& 1400
& $3.0\times10^{19}$
& 2700
& 1.05
& 25
& 2.05
& 107
& 2.86 \\
RRCAT \cite{Tayyab_2019}
& 150-TW Ti:Sa
& 2.5
& $1.0\times10^{20}$
& 25
& 0.8
& 1
& 3.02
& 2.74
& 2.56 \\
LULI \cite{Baccou_Depierreux_2015}
& ELFIE
& 12
& $1.0\times10^{19}$
& 350
& 1.056
& 20
& 1.03
& 5.84
& 2.25 \\
LULI \cite{labaune_fusion_2013}
& Pico2000
& 20
& $6.0\times10^{18}$
& 1000
& 0.53
& 20
& 0.25
& 21.4
& 1.24 \\
\bottomrule
\end{tabular}
\normalsize
\end{table}

\section{Discussion}
The present analytical framework enables the extraction of the low-energy astrophysical $S$-factor. In laser-driven experiments, the measured fusion yield can be expressed as $N_{\mathrm{fus}} = \nu\, n_i\, n_t\, \langle\sigma v\rangle\, V\, \tau$, where $\nu$ is the number of reaction products per fusion event, $n_i$ is the density of the incident ion beam in the catcher target, $n_t$ is the catcher-target density, $V$ is the interaction volume, and $\tau$ is the interaction time. For a given experimental configuration in which $N_{\mathrm{fus}}$, $n_i$, $n_t$, $V$, and $\tau$ are measured or independently constrained, the reactivity $\langle\sigma v\rangle$ can be inferred. Substituting this inferred reactivity into Eq.\,(\ref{eq_an_reac}), which relates $\langle\sigma v\rangle$ to $S(E_0)$, allows one to determine $S(E_0)$ from the experimentally measured fusion yield. Since the effective energy $E_0$ depends on $T_e$, operating in a low-$T_e$ regime gives access to lower-energy $S(E_0)$ values, which are essential for predicting elemental abundances in astrophysical environments.

More generally, compared with conventional accelerator experiments, laser-driven nuclear reactions extract the astrophysical $S$-factor indirectly from the measured fusion yield. In this case, the yield is determined by the convolution of the reaction cross section with the broad ion-energy distribution, and the corresponding kernel function contains the energy dependence of the $S$-factor. In contrast, accelerator experiments generally determine the reaction cross section at a well-defined energy using a quasi-monoenergetic ion beam. One advantage of the laser-driven approach is that a broad reaction-energy range can be probed within a single laser shot. This may provide access to low-energy reaction information in cases where direct accelerator measurements are limited by very low yields or experimental backgrounds. In particular, when a reaction contains a narrow resonance in the low-energy region, the broad laser-driven ion spectrum can overlap with the resonant energy range without requiring the incident beam energy to be tuned precisely to the resonance.

For a narrow resonance, as an example, the above formalism can be extended to include resonant reactions.
In general, the total reactivity is given by the non-resonant component and all relevant resonant contributions, $\left\langle \sigma v \right\rangle_{\rm total} = \left\langle \sigma v \right\rangle_{NR} + \sum\left\langle \sigma v \right\rangle_{R}$.
For a narrow resonance at $E=E_R$, the reaction cross section between the initial channel $i$ 
($a + A \rightarrow R^*$) 
and the final channel $f$ ($R^* \rightarrow b + B$) 
is described by the Breit--Wigner formula,
\begin{equation}
\sigma_{\mathrm{R}}^{(i,f)}(E)
= \frac{\pi}{k_i^2}\,
\omega\,\frac{\Gamma_i \Gamma_f}{(E - E_R)^2 + (\Gamma/2)^2},
\label{BR}
\end{equation}
where $k_i = \sqrt{2\mu_i E}/\hbar$ is the wave number in the entrance channel, 
$\omega = (2J_R + 1)/[(2J_A + 1)(2J_a + 1)]$ is the statistical factor, 
and $\Gamma_i$, $\Gamma_f$, and $\Gamma$ are the entrance partial, exit partial, and total widths, respectively.

In the standard thermal case, a narrow resonance effectively confines the integration to $E = E_R$, 
so that the reactivity integral reduces to the cross section integration of Eq.\,(\ref{BR}), 
which yields $(2\pi^2/k_i^2)\,\omega\,\gamma$~\cite{XU201361}, 
where $\omega\gamma = \Gamma_i \Gamma_f / \Gamma$ is the resonance strength that can be determined experimentally. 
Similarly, substituting Eq.\,(\ref{BR}) into Eq.\,(\ref{eq:rate_beam}), the integrand becomes sharply peaked at $E_i \simeq E_R$, 
and the integral can be evaluated using the narrow-resonance approximation, yielding
\begin{equation}
\langle \sigma v \rangle_{\mathrm{R}}
\simeq \sqrt{\frac{2}{m_i}}\,
\sqrt{E_R}\,
f_{i,\mathrm{ss}}(E_R)\,
\left( \frac{2\pi^2}{k_i^2}\right)\,\omega\,\gamma.
\label{eq:rate_res}
\end{equation}
This result can also be obtained from Eq.\,(\ref{eq:rate_beam}) by substituting the cross section with $\sigma(E) = (2 \pi^2 / k_i^2) \omega \gamma \delta(E_i-E_R)$ function. Consequently, once the resonance parameters, resonance energy $E_R$ and partial widths $\Gamma_i$ and $\Gamma_f$, are specified, 
Eq.\,(\ref{eq:rate_res}) provides a direct means to evaluate the fusion reactivity for resonant reactions within the same theoretical framework developed here for non-resonant reactions. For reactions involving broad resonances, numerical calculations are required. For example, in the case of the $p + ^{11}\mathrm{B}$ reaction, such an analysis has been reported in Ref.\,\cite{Hwang_2025}.

Despite these advantages, several challenges must be addressed to determine the astrophysical $S$-factor accurately from laser-driven nuclear reactions. Compared with conventional accelerator measurements, additional uncertainties arise from shot-to-shot variations of the ion spectrum, the total number of accelerated ions, target conditions and thickness effects, possible electron-screening effects, detector response, and the model dependence associated with reconstructing the ion distribution and the effective reaction-energy window. A quantitative assessment of these uncertainties is beyond the scope of the present work, as it requires experimentally constrained diagnostics of the ion spectrum, ion number, target properties, and fusion yield. In particular, a reliable determination of the ion spectrum and its interaction with the catcher target requires more sophisticated analyses incorporating detailed target-plasma dynamics. At the present stage, the laser-driven approach should therefore be regarded as complementary to, rather than a replacement for, precision monoenergetic accelerator measurements. Its absolute accuracy is expected to be limited by the experimentally constrained uncertainties in the ion spectrum, ion number, target response, and detector calibration. Electron screening may also become relevant depending on the target density, temperature, ionization state, and reaction channel, especially for lower effective reaction energies. PIC and Monte Carlo simulations will be necessary to quantify and reduce the associated model dependence, and such developments will be studied in future work.

\section{Conclusion}
In conclusion, we have developed an analytical framework for laser-driven fusion reactivities that incorporates the non-Maxwellian ion distributions generated by TNSA acceleration. Using a self-similar plasma-expansion model, we derived a closed-form expression for the effective reaction energy $E_0$, analogous to the Gamow peak in thermal plasmas. Assuming a slowly varying $S(E)$, the reactivity $\langle\sigma v\rangle$ can be expressed in terms of the modified Bessel function $K_0$, revealing an optimal electron temperature $T_{e,\max}$ that maximizes the fusion rate and implies an optimal laser condition for pitcher–catcher experiments. The formulation was further extended to narrow resonances, allowing consistent evaluation of resonant contributions once $E_R$ and the partial widths are given.

This analytical framework provides a predictive tool for estimating fusion reactivities and effective energy windows under specified target, beam, and laser parameters. It further enables the extraction of the low-energy astrophysical $S$-factor from measured fusion yields when operated in a low-$T_e$ regime, thereby linking laser-driven fusion studies to nuclear astrophysics. Future developments may incorporate experimental uncertainties associated with the ion spectrum, ion number, and target conditions, as well as additional physics relevant to real experiments, including non-Maxwellian electron distributions \cite{PhysRevLett.40.1652, 10.1063/1.862751, 10.1063/1.864389}, time-dependent $T_e$ \cite{Gurevich1981JETP}, multidimensional effects \cite{PhysRevLett.85.2945, Wilks2001}, magnetic fields \cite{PhysRevLett.84.670, PhysRevLett.86.3562}, finite initial ion-density gradients \cite{Wilks2001, PhysRevLett.86.1769, PhysRevSTAB.5.061301}, and ionization dynamics \cite{PhysRevLett.89.085002}.
\\

\ack
E.H. and M.-K.C. are supported by the Basic Science Research Program of the National Research Foundation of Korea (NRF) under Grant Nos. RS-2021-NR060129, RS-2024-00460031, and RS-2025-16071941. H.K was supported by the Deutsche Forschungsgemeinschaft (DFG, German Research Foundation) - Project-ID 279384907 - SFB 1245.

\section*{References}
\bibliographystyle{iopart-num} 
\bibliography{ref}

\end{document}